\documentclass[preprint,aps,tightenlines,superscriptaddress,showpacs,showkeys]{revtex4}
\usepackage{epsfig}
\usepackage{amsmath}
\newcommand{\vi}[1]{\mbox{\boldmath $#1$}}
\newcommand{\vis}[1]{\mbox{\boldmath ${\scriptstyle #1}$}}
\begin{document}
\title{Phase-shift calculation using continuum-discretized states} 
\author{Y. Suzuki}
\affiliation{Department of Physics, and Graduate School of 
Science and Technology, Niigata University, Niigata
950-2181, Japan} 
\email{suzuki@nt.sc.niigata-u.ac.jp}
\author{W. Horiuchi}
\affiliation{Graduate School of Science and Technology, Niigata
University, Niigata 950-2181, Japan}
\email{horiuchi@nt.sc.niigata-u.ac.jp}
\author{K. Arai}
\affiliation{Division of General Education, Nagaoka National 
College of Technology, Nagaoka 940-8532, Japan}
\email{arai@nagaoka-ct.ac.jp}

\begin{abstract}
We present a method for calculating scattering phase shifts which 
utilizes continuum-discretized states obtained in a bound-state type 
calculation. The wrong asymptotic behavior of the discretized state 
is remedied by means of the Green's function formalism. Test examples confirm  
the accuracy of the method. The $\alpha+n$ scattering is described 
using realistic nucleon-nucleon potentials. The $3/2^-$ and $1/2^-$ 
phase shifts obtained in a single-channel calculation 
are too small in comparison with experiment. The $1/2^+$ phase 
shifts are 
in reasonable agreement with experiment, and gain contributions 
both from 
the tensor and central components of the nucleon-nucleon potential.    
 
\end{abstract}
\maketitle
\draft

\section{Introduction}

Accurate solutions for discrete states of few-nucleon systems 
interacting via a realistic potential 
have been obtained with various sophisticated 
methods~\cite{benchmark}. The most significant 
merit of these methods is that the interaction 
employed is tested strictly in comparison with experiment 
and thus the nuclear structure can be 
understood without {\it ad hoc} assumptions, and some important 
ingredients such as a three-body force are revealed. 
In contrast to the bound-state problem, a microscopic 
description of reactions using a realistic potential has been 
hampered by the difficulties related to continuum states 
as well as scattering boundary conditions. 
Some progress has recently been made towards the application of an 
{\it ab initio} approach to the problem of 
scattering and resonances~\cite{nollet,quaglioni}. The study 
in this direction should be further pursued as it may 
provide more detailed information on 
the characteristics of the interaction as functions of partial 
waves and incident energies and on the dynamics of the participating 
nuclei. 

Because of their apparently different natures, continuum and 
discrete states are obtained separately using different methods. 
Since, however, both of them are solutions of the same 
Schr{\"o}dinger equation, it would be nice if they could be 
obtained on an equal footing, namely if the 
correlation of the constituent particles could be included 
in the scattering problems as well as in the bound-state problems. 
Several methods have in fact been proposed using square-integrable 
(${\cal L}^2$) basis functions for continuum problems in atomic and nuclear 
physics~\cite{harris,hazi,ivanov,savukov,mitroy,kruppa}.  

The purpose of this article is to show a simple approach to 
the scattering problem using the technique for discrete 
states, particularly developed for a basis expansion method. 
The present approach has some similarity to that of  
Ref.~\cite{ivanov}, where the stochastic variational
method~\cite{kukulin,svm} is used in conjunction with stabilization
ideas to calculate the low energy phase-shifts  
for positronium-atom scattering. The basis set for describing the 
configuration space of the interaction region is 
spanned by the stochastic 
variational method. In Ref.~\cite{ivanov}, the phase shift is 
extracted by fitting the wave 
function in the scattering region to 
$\sin(kr+\delta)$. 
We instead use the Green's function approach to describe the wave
function in the scattering region. 

The basic quantity in the scattering of nuclei is the overlap integral 
between the product of the two internal wave functions of the nuclei 
and the scattering 
wave function of the composite system. This overlap integral is called 
a spectroscopic amplitude (SA) in this article. The phase shift is
determined from the asymptotic behavior of the SA. This type of overlap 
integral was studied long time ago for extracting spectroscopic
information from single-particle transfer reactions~\cite{berggren,pinkston}.  
In Sec.~\ref{formalism}, we discuss the equation of motion which the SA 
satisfies, and we derive a formula to calculate the phase shift. The 
accuracy of the present approach is tested in Secs.~\ref{np.scat} and 
\ref{alpha-n.effective}. Section~\ref{alpha-n.real} presents an 
application to $\alpha+n$ scattering using a realistic nucleon-nucleon 
potential model. A brief summary is given in Sec.~\ref{summary}.
The calculation of the SA is explained in the Appendix.

\section{Formalism}
\label{formalism}

Let $c$ stand for a channel including a pair of nuclei 
denoted $\alpha_1$ and $\alpha_2$ whose normalized wave functions, 
$\psi_{I_1}(\alpha_1)$ and 
$\psi_{I_2}(\alpha_2)$, are assumed to be given. 
The wave function of the total system with energy $E$ generally 
takes the form 
\begin{equation}
\Psi_{JM}=\sum_c\Psi_{cJM}+\sum_{\gamma} {\Xi}_{\gamma JM}.
\label{cdcc}
\end{equation}
The first term on the right side stands for configurations 
for the elastic channel as well as other channels, whereas  
the second term represents the configurations that are needed to 
take account of the effects of resonances, distorted states, etc., 
not included in the first term.  
The term $ {\Xi}_{\gamma JM}$ is thus assumed to represent the  
configurations in which all the nucleons are confined in the 
interaction region. It is assumed that $\Psi_{JM}$ satisfies 
the antisymmetry requirement for the exchange of the nucleons. 

A key in the present method is the SA 
\begin{equation}
y(r)=\langle \Phi_{cJM}(r) | \Psi_{JM} \rangle,
\label{spectr.amp}
\end{equation}
where $\Phi_{cJM}(r)$ is a test function for the channel $c$
\begin{equation}
\Phi_{cJM}(r)=\left[
[\psi_{I_1}(\alpha_1)\psi_{I_2}(\alpha_2)]_I 
Y_{\ell}(\hat{\vi r}_c)\right]_{JM}\frac{\delta(r_c-r)}{r_cr}.
\label{test.fn}
\end{equation} 
The coordinate ${\vi r}_c$ denotes the relative distance 
vector of the two nuclei. 
The angular momenta $I_1$ and $I_2$ of the two nuclei 
are coupled to the channel spin $I$, which is then coupled with 
the orbital angular momentum ${\ell}$ of the relative motion to 
the total angular momentum $JM$. In the test 
function~(\ref{test.fn}), the two nuclei are pinned down 
at the relative distance specified by $r$.

The phase shift for the scattering of nuclei $\alpha_1$ and 
$\alpha_2$ is calculated from the asymptotic 
behavior of $y(r)$.  When $\Psi_{JM}$ is obtained 
in a bound-state approximation, the SA calculated from it 
has ill behavior at large distances, and hence the 
asymptotics of $y(r)$ is usually 
not good enough to enable one to calculate the phase shift 
accurately. To resolve this problem, we derive an equation of 
motion which $y(r)$ calculated using the exact $\Psi_{JM}$ should 
satisfy. For this purpose we start from the equation 
\begin{equation}
\langle \Phi_{cJM}(r) |H| \Psi_{JM}\rangle=E 
\langle \Phi_{cJM}(r) |\Psi_{JM}\rangle.
\label{sch.eq}
\end{equation}
The Hamiltonian of the system can be decomposed into
\begin{equation}
H=H_{\alpha_1}+H_{\alpha_2}+T_c +V_c,
\label{hamiltonian}
\end{equation}
where $H_{\alpha_1}$ and $H_{\alpha_2}$ are, respectively, 
the internal Hamiltonians of nuclei $\alpha_1$ and $\alpha_2$, 
\begin{equation}
T_c=-\frac{\hbar^2}{2\mu_{c}}\frac{\partial^2}
{\partial {\vi r}_{c}^2 }
\end{equation} 
is the kinetic energy of the relative motion between them, and
\begin{equation}
V_c=\sum_{i\in \alpha_1,\, j\in \alpha_2}v_{ij}
\end{equation} 
is the interaction acting between the two nuclei. 
If the Hamiltonian contains a three-body force, the potential 
$V_c$ acting between the nuclei should include the following 
terms of 
the three-body potentials, 
$\sum_{({i\in \alpha_1})< (j<k\in \alpha_2)}v_{ijk}$+
$\sum_{({i<j\in \alpha_1})< (k\in \alpha_2)}v_{ijk}$.

Substituting Eq.~(\ref{hamiltonian}) into Eq.~(\ref{sch.eq}) 
and introducing 
a local potential $U_c(r)$ which acts between nuclei $\alpha_1$ 
and $\alpha_2$ makes it 
possible to transform Eq.~(\ref{sch.eq}) to the following 
inhomogeneous differential equation for $y(r)$ (the 
subscript $c$ in $U_c$ and $\mu_c$ is suppressed hereafter)
\begin{equation}
\left[\frac{d^2}{dr^2}+
\frac{2}{r}\frac{d}{dr}-\frac{\ell(\ell+1)}{r^2} 
-\frac{2\mu}{\hbar^2}U(r)+k^2 \right] y(r)
 =\frac{2\mu}{\hbar^2}[z(r)+w(r)],
\label{eq.y-z}
\end{equation}
where $k$=$\sqrt{{2\mu}(E-E_{\alpha_1}-E_{\alpha_2})/{\hbar^2}}$ 
with $E_{\alpha_i}=\langle
\psi_{I_iM_i}|H_{\alpha_i}|\psi_{I_iM_i}\rangle $ 
is the wave number for the relative motion, and the functions 
$z(r)$ and $w(r)$ are, respectively, defined by 
\begin{equation}
z(r)=\langle \Phi_{cJM}(r)\mid V_c-U \mid \Psi_{JM}\rangle,
\label{zfunc}
\end{equation} 
and 
\begin{equation}
w(r)=\langle \Phi_{cJM}(r)\mid H_{\alpha_1}-E_{\alpha_1}
+H_{\alpha_2}-E_{\alpha_2} \mid \Psi_{JM}\rangle.
\end{equation}
The function $w(r)$ vanishes if 
$\psi_{I_iM_i}$ are the eigenfunctions of $H_{\alpha_i}$, 
that is, $H_{\alpha_i}\psi_{I_iM_i}$=$E_{\alpha_i}\psi_{I_iM_i}$.  

Equation~(\ref{eq.y-z}) is apparently equivalent to 
Eq.~(\ref{sch.eq}), which is the 
Schr\"{o}dinger equation projected to the space 
spanned by the test function $\Phi_{cJM}(r)$. The equivalence 
does hold for an arbitrary choice of $U(r)$. 
Let $v(r)$ and $h(r)$ denote, respectively, the regular 
and irregular solutions of the homogeneous equation 
with $z(r)+w(r)$ being set to zero in 
Eq.~(\ref{eq.y-z}). They are defined to satisfy the Wronskian 
relation, 
$W(v,h)(r)\equiv v(r)h'(r)-v'(r)h(r)$=$1/(kr^2)$. 
A general solution of Eq.~(\ref{eq.y-z}) that is regular at 
$r$=0 and has the asymptotic behavior appropriate to the 
scattering solution reads 
\begin{equation}
y(r)=\lambda v(r)
+\frac{2\mu}{\hbar^2}\int_0^{\infty}G(r,r')[z(r')+w(r')]r'^2dr',
\label{y.formal}
\end{equation}
where $\lambda$ is a constant. Here the 
Green's function $G$ which is a solution of the following equation
\begin{equation}
 \left[\frac{d^2}{dr^2}+\frac{2}{r}\frac{d}{dr}
-\frac{\ell(\ell+1)}{r^2}  -
\frac{2\mu}{\hbar^2}U(r)+k^2\right]G(r,r')=
\frac{1}{rr'}\delta(r-r'),
\label{greenf}
\end{equation}
is given by~\cite{suzuki78}
\begin{equation}
G(r,r')=\begin{cases}
k v(r)h(r')  & \text{$r \le r'$} \\
k h(r)v(r')  & \text{$ r \ge r' $}. 
\end{cases}
\label{ggfn}
\end{equation}

By expressing the integral in Eq.~(\ref{y.formal}) as 
\begin{equation}
\int_0^{\infty}G(r,r')[z(r')+w(r')]r'^2dr'=k[p(r)h(r)+q(r)v(r)]
\end{equation}
with
\begin{equation}
p(r)=\int_0^r v(r')[z(r')+w(r')]r'^2dr',\ \ \ \ \ \ 
q(r)=\int_r^{\infty}h(r')[z(r')+w(r')]r'^2dr',
\end{equation}
the SA~(\ref{y.formal}) takes the form 
\begin{equation}
y(r)=\left[\lambda +\frac{2\mu k}{\hbar^2}q(r)\right]v(r)
+\frac{2\mu k}{\hbar^2}p(r)h(r).
\label{y.exact}
\end{equation}
Taking the asymptotics of this $y(r)$ determines the phase shift 
$\delta_{\ell}$ as 
\begin{equation}
\tan \delta_{\ell}=\tan \delta_{\ell}^{(0)}-
\frac{2\mu k}{\hbar^2 \lambda}p(\infty),
\label{ps.formula}
\end{equation}
where $\delta_{\ell}^{(0)}$ is the phase shift corresponding 
to the potential scattering by $U(r)$. 
Let $y(r)$ of Eq.~(\ref{y.exact}) (or Eq.~(\ref{y.formal})) 
be called SAGF (SA solved with 
the Green's function). 

Equation~(\ref{ps.formula}) shows that we can obtain 
the phase shift accurately if $\lambda$ and $p(\infty)$ are 
known to high accuracy. The value of $p(\infty)$ consists 
of two terms, one involving 
the function $z(r')$ and the other involving the function $w(r')$. 
As was already mentioned, $w(r')$ vanishes if $\psi_{I_iM_i}$ are 
the eigenfunctions of the internal Hamiltonian $H_{\alpha_i}$. 
As we will see later, even though $\psi_{I_iM_i}$ are not completely 
identical with the eigenfunctions, the magnitude of $w(r')$ turns out to 
be much smaller than that of $z(r')$. 
The function $z(r')$ consists of the sum of various pieces 
of $V_c$ such as central, tensor and spin-orbit forces, and thus  
$\tan \delta_{\ell}$ can be decomposed into the contributions 
of those terms. 

In a practical calculation of a phase shift, we have only an 
approximate solution for $\Psi_{JM}$, and because of this 
$y(r)$ as well as $z(r)$ and $w(r)$ are all approximately 
evaluated. The value of $\lambda$ is determined 
by comparing the two SAs, Eqs.~(\ref{spectr.amp}) 
and (\ref{y.exact}), as will be discussed later. Evaluating 
$p(\infty)$ to a good approximation depends on a choice of $U(r)$. 
We assume that $U$ is chosen in such a way that $V_c$ approaches 
$U$ for large $r$, namely $V_c-U$ is an operator that is non-zero 
only in the interaction region. If $U$ 
is chosen to satisfy this condition, $z(r)$ can be accurate even 
though $\Psi_{JM}$ does not have a correct tail in the 
region where $V_c-U$ is negligibly small. The function $w(r)$ 
is accurate as well in the interaction region. Then we may 
assume that both $p(r)$ 
and $q(r)$ can be evaluated fairly accurately provided $U$ is 
suitably chosen. 

The functions 
$z(r)$ and $w(r)$ (and $y(r)$ of Eq.~(\ref{spectr.amp})) are 
calculated using a code for bound-state calculations in so 
far as the Dirac $\delta$-function in the 
test function~(\ref{test.fn}) is approximated as 
\begin{equation}
\frac{\delta(r_c-r)}{r_cr}\approx \sum_{\nu}f_{\nu}(r)f_{\nu}(r_c),
\label{delta.exp}
\end{equation}
where $\{f_{\nu}\}$ is 
an ${\cal L}^2$ `pseudo-complete' set with 
$\langle f_{\nu}\mid f_{\nu'}\rangle=\delta_{\nu \nu'}$~\cite{beck}. 
The expansion like Eq.~(\ref{delta.exp}) is used also 
in Ref.~\cite{quaglioni} in the basis of harmonic-oscillator 
functions. 
Because a more precise evaluation of $y(r)$ is in 
general desirable, we show in the Appendix an analytical method 
to calculate the SA~(\ref{spectr.amp}) for the correlated Gaussian 
basis functions~\cite{svm,fbs} which are employed in this article. 
The accuracy of the series expansion (\ref{delta.exp}) in the
calculation of functions of the type of $z(r)$ 
was discussed in Ref.~\cite{lovas} 
by comparing it to the exact calculation for the $\alpha+t$
system where the two fragments are assumed as the lowest shell-model 
states with a common oscillator parameter. 

An equation similar to Eq.~(\ref{eq.y-z}) 
was proposed to improve the overlap integrals which appear in 
nucleon transfer reactions or virtual nucleon 
decay~\cite{kawai,timofeyuk} and $\alpha$ 
decay~\cite{horiuchi73}. 
To our knowledge, no one has yet used Eq.~(\ref{eq.y-z})  
to obtain the phase shift in a microscopic calculation.

In a microscopic reaction theory such as 
the resonating group method (RGM)~\cite{rgm}, $\Psi_{cJM}$ of 
Eq.~(\ref{cdcc}) is expressed as 
\begin{equation}
\Psi_{cJM}=\int_0^{\infty} u_{\ell}(r){\cal A}\Phi_{cJM}(r)r^2dr,
\label{rgm.wf}
\end{equation}
where $\cal A$ is the internucleus antisymmetrizer, and the 
phase shift for the elastic scattering 
is calculated from an integro-differential equation for 
$u_{\ell}(r)$~\cite{baye}. 
We stress that $y(r)$ is used in the present approach 
instead of $u_{\ell}(r)$. The function $y(r)$ is  
subject to the simple differential equation, and it is  
determined uniquely regardless of whether or not Pauli-forbidden 
state exist, which is in contrast to the case of $u_{\ell}(r)$. 

The accuracy of the phase-shift calculation crucially depends on 
how accurately $\lambda$ is determined. We tested two ways to 
determine $\lambda$. 
The first is to fit the SAGF~(\ref{y.exact}) which is a function of 
$\lambda$ to the 
SA~(\ref{spectr.amp}) with the least squares method in the 
interval $[r_0,r_1]$ where the SA~(\ref{spectr.amp}) is 
expected to be accurately obtained:
\begin{equation}
{\rm minimize\  over\  \lambda}: \sum_{i\,(r_0 \le r_i \le r_1)}
[y^{\rm SAGF}(r_i)-y^{\rm SA}(r_i)]^2.
\label{method.1}
\end{equation}  
We found that $\lambda$ determined in this way remains 
virtually unchanged within moderate choices of the interval.  
The second is to calculate the Wronskian $W(y,h)(r)$ 
using both $y^{\rm SA}(r)$ and $y^{\rm SAGF}(r)$. The latter reads 
$W(y^{\rm SAGF},h)(r)=(\lambda/kr^2)+(2\mu/\hbar^2r^2)q(r)$. 
Equating the 
two Wronskians leads to the following expression for $\lambda$: 
\begin{equation}
\lambda(r)=kr^2W(y^{\rm SA},h)(r)-\frac{2\mu k}{\hbar^2}q(r)
=kW(ry^{\rm SA},rh)(r)-\frac{2\mu k}{\hbar^2}q(r),
\label{det.lambda}
\end{equation}
which usually becomes $r$-dependent because $y^{\rm SA}(r)$ is only 
approximately equal to $y^{\rm SAGF}(r)$.
The least squares fitting to this $\lambda(r)$ in the interval 
$[r_0,r_1]$,
\begin{equation}
{\rm minimize\  over\  \lambda}: \sum_{i\,(r_0 \le r_i \le r_1)}
[\lambda(r_i)-\lambda]^2,
\label{method.2}
\end{equation}
yields again 
a stable $\lambda$ value, which is in good agreement with that 
determined by the first method. We use the first one mostly in what 
follows because it requires no differentiation of $y(r)$. 
Interestingly, unlike the $R$-matrix theory~\cite{R.matrix}, 
our phase-shift calculation requires no channel radius.

As shown in Ref.~\cite{suzuki78}, the 
construction of the Green's function is easy even for
coupled-channel problems if $U$ is local. 
It is thus noted that the present approach can be straightforwardly 
extended to the scattering including coupled-channels. 

\section{Examples}

\subsection{$^3S_1$ $n+p$ scattering}
\label{np.scat}

A first example to test the present approach is the 
$^3S_1$ $n+p$ scattering phase shift calculated 
with the Minnesota potential~\cite{mn}. This is just a potential 
scattering of the two particles, and a numerically exact 
phase shift can easily be obtained.  
Diagonalizing the $n+p$ Hamiltonian with the Minnesota potential $v(r)$ 
in appropriate ${\cal L}^2$ basis functions produces, 
besides the deuteron ground state, continuum discretized states 
and corresponding energies $E$. 
The phase shifts are calculated using these states. The ${\cal L}^2$ 
basis functions used for the $S$ wave are 
Gaussians, $\exp(-\frac{1}{2}\beta r^2)$ with different 
falloff parameters $\beta$, where $\beta$ is real or complex 
with Re\,$\beta >$\,0~\cite{hiyama}. In this potential 
problem, $y^{\rm SA}(r)$ for the discretized energy $E$ 
takes the form 
\begin{equation}
y^{\rm SA}(r)=\sum_{i=1}^K C_i(E)\, {\rm e}^{-\frac{1}{2}\beta_i r^2},
\end{equation}
and the function $z(r)$ reduces to $v(r)y^{\rm SA}(r)$, and 
$w(r)$ vanishes. The potential $U(r)$ is set to zero.

\begin{table}[t]
\caption{Comparison of the $^{3}S_1$ phase shifts, given in degrees, 
of $n$+$p$ scattering between Numerov and SAGF methods. The $\lambda$ 
value is determined using Eq.~(\ref{method.1}) (Method 1) or  
(\ref{method.2}) (Method 2) with the use of different intervals 
$[r_0,r_1]$ (fm). The Minnesota 
potential~\cite{mn} is used.}
\begin{tabular}{cccccccccccc}
\hline\hline
$E$&&&Numerov&&&\multicolumn{3}{c}{Method 1}&&&\multicolumn{1}{c}
{Method 2}\\
\cline{7-9}\cline{11-12}
[MeV]  &&&  &&&$[0,5]$&$[1,6]$&$[2,6]$&&&$[2,6]$\\
\hline
0.4986&&&147.7 &&&147.7 &147.8 &147.7&&&147.7\\
1.959 &&&123.2 &&&123.3 &123.2 &123.2&&&123.3\\
4.395 &&&105.3 &&&105.2 &105.3 &105.3&&&105.3\\
7.948 &&&91.2  &&&91.4  &91.2  &91.2 &&&91.1\\
12.87 &&&79.2  &&&79.0  &79.2  &79.2 &&&79.4\\
19.54 &&&68.5  &&&68.7  &68.5  &68.5 &&&68.3\\
28.49 &&&58.5  &&&58.2  &58.6  &58.6 &&&58.8\\
40.42 &&&49.3  &&&49.6  &49.3  &49.3 &&&49.1\\ 
56.28 &&&40.8  &&&40.4  &40.8  &40.8 &&&40.8\\
77.31 &&&33.2  &&&33.6  &33.1  &33.2 &&&33.3\\
\hline\hline
\end{tabular}
\label{comp.np3s1}
\end{table}

\begin{figure}[b]
\epsfig{file=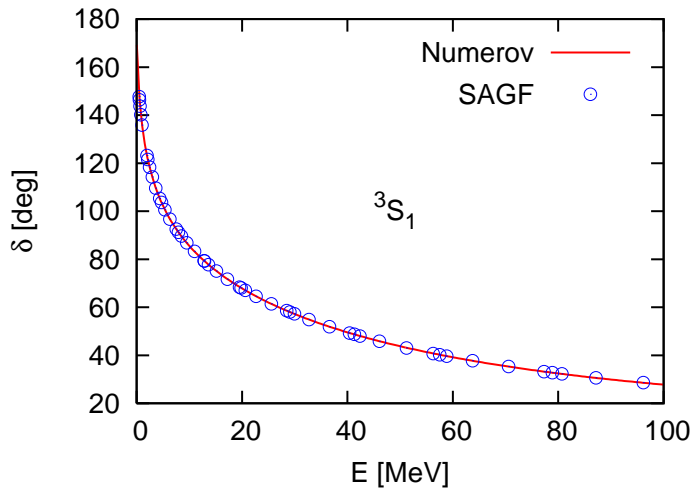,scale=1.5}
\caption{Comparison of the $^{3}S_1$ phase shifts of $n$+$p$ scattering 
between SAGF and Numerov methods. The Minnesota 
potential~\cite{mn} is used.} 
\label{np3s1}
\end{figure}

Table~\ref{comp.np3s1} compares the phase shifts 
calculated using Eq.~(\ref{ps.formula}) 
with those obtained with the Numerov method, which is virtually 
exact. The comparison shows that the present method produces 
very stable phase shifts which are rather insensitive to the 
choice of the method of determining $\lambda$ as well as the 
interval used for the minimization of the error. 
To generate the discretized states for different energies, we 
repeated the calculation by changing $K$ and the set of $(\beta_1,
\beta_2,\ldots, \beta_K)$. Figure~\ref{np3s1} 
displays the phase shifts obtained in this way.  
The phase shifts obtained with our method almost
perfectly agree with those of the Numerov method 
in a wide range of incident energies. 

\subsection{$\alpha+n$ scattering with an effective 
nucleon-nucleon potential}
\label{alpha-n.effective}

In this and the 
following subsections we calculate the $S$- and $P$-wave 
phase shifts of the 
$\alpha+n$ scattering in a microscopic approach. 
The nucleon-nucleon interaction employed in this subsection 
is the Minnesota 
potential~\cite{mn} which consists only of the central and 
spin-orbit 
potentials. The $u$ parameter of the central potential is set 
equal to 0.98, and the spin-orbit potential adopted is 
$-591.1\,{\rm e}^{-3r^2}{\vi \ell}\cdot{\vi \sigma}$ in MeV (set IV 
of Reichstein and Tang). This potential is known to reproduce 
the empirical $\alpha+n$ phase shifts of Refs.~\cite{bond,stamm}. 
The phase-shift difference between the $3/2^-$ 
and $1/2^-$ states is particularly well reproduced by 
this spin-orbit 
potential, and in this 
sense the Minnesota potential can be regarded as an effective 
potential. 

Only a single $\alpha+n$ channel is 
included in the phase-shift calculation. The wave function of 
the $\alpha$ particle, $\psi_0(\alpha)$, is obtained by 
diagonalizing 
the Hamiltonian $H_{\alpha}$ in a basis of a number of Gaussians. 
The 
binding energy with the Coulomb potential being included is 
$29.90$\,MeV and the root-mean-square matter radius is 1.41\,fm.

The relative motion function $u_{\ell}(r)$ in Eq.~(\ref{rgm.wf}) 
is taken as a combination of Gaussians:
\begin{equation}
u_{\ell}(r)=\sum_{i=1}^K C_i \,r^{\ell}\,
{\rm e}^{-\frac{1}{2}\beta_i r^2}.
\label{uexp}
\end{equation}
We have calculated the phase shifts using two different methods, 
the microscopic $R$-matrix theory~\cite{micro.R-matrix} and 
the present method. The 
accuracy of the $R$-matrix theory is well tested, and it is 
considered to produce virtually 
exact results to which the phase shifts of SAGF are to be 
compared. In the SAGF calculation the parameters 
$b_i=1/\sqrt{\beta_i}$ with real $\beta_i$ are chosen 
to form a geometric 
progression to cover $0< b_i \lesssim 10$\,fm. 
The number $K$ of the basis functions is about 10-15.
The Gaussian basis used in 
Eq.~(\ref{uexp}) can also be employed to construct the 
pseudo-complete set in Eq.~(\ref{delta.exp}). We obtained 
all the matrix elements needed in SAGF using the method developed 
in Ref.~\cite{fbs}. See also the Appendix for the calculation of 
the SA. 

\begin{figure}[b]
\epsfig{file=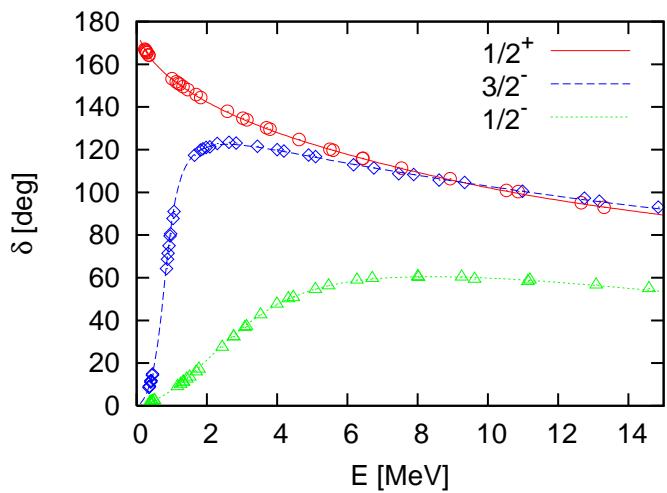,scale=1.5}
\caption{Comparison of the phase shifts of $\alpha$+$n$ scattering 
between the microscopic $R$-matrix and SAGF methods. Solid, 
dashed and dotted lines are the phase shifts calculated with the 
$R$-matrix theory, while symbols are those by SAGF. 
The Minnesota potential~\cite{mn} is used.} 
\label{alpha.nMN}
\end{figure}

Figure~\ref{alpha.nMN} compares the SAGF phase shifts with the 
$R$-matrix phase shifts. The value of $\lambda$ is determined 
by the first method using the interval $[1,5]$\,(fm). 
The agreement is excellent. 
Both of $z(r)$ and $w(r)$ are included in the SAGF calculation. 
We switched off the function $w(r)$ and found that the 
phase-shift change is negligible: The largest change of 
a few percent occurs in the 
resonance region at around 1\,MeV of the $3/2^{-}$ phase shifts.  
Except for this case the phase-shift change is smaller by one 
order of magnitude. Thus we may safely neglect the contribution of $w(r)$.

\begin{figure}[t]
\epsfig{file=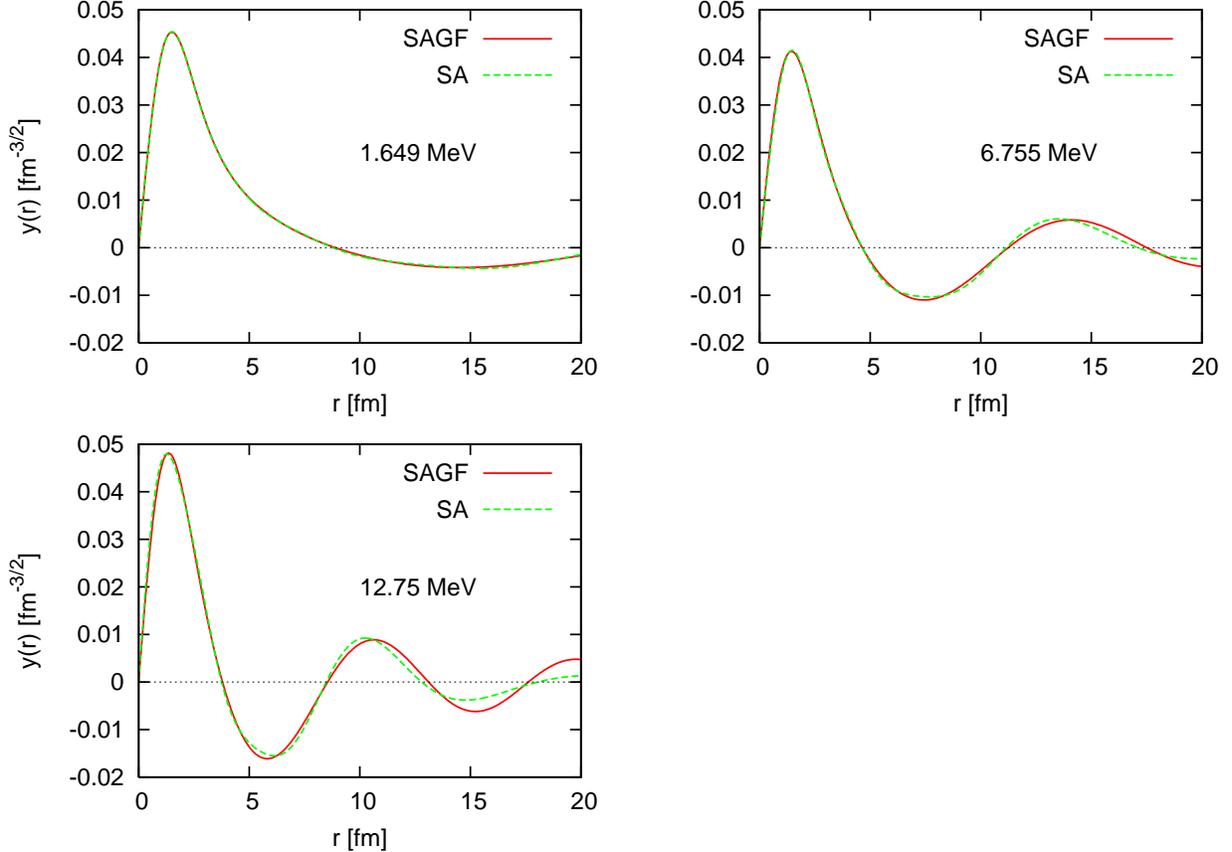,scale=1.3}
\caption{Comparison between the SA and the SAGF for the
 $\alpha$+$n$ scattering ($3/2^-$) at three different energies. The Minnesota
 potential~\cite{mn} is used.}
\label{comp.sa.sagf}
\end{figure}

It is interesting to see how much the SAGF changes from the
SA. Figure~\ref{comp.sa.sagf} compares the two SAs for the $3/2^-$ state 
at three different energies. In each case, the SA curve agrees very well 
with the corresponding SAGF in the region of $r<5$\,fm, which 
indicates that the expansion~(\ref{uexp}) is good enough to 
describe to a good approximation the $\alpha$-$n$ relative motion function 
in the interaction region. The peak position and 
the amplitude of the SA curve begin to deviate from those of the 
SAGF curve for $r>5$\,fm. The deviation becomes larger as the 
energy increases. 

\subsection{$\alpha+n$ scattering with realistic 
nucleon-nucleon potentials}
\label{alpha-n.real}

In this subsection we take the same model as in the previous subsection
but employ the realistic nucleon-nucleon potentials of AV8 type, 
AV8$^{\prime}$~\cite{av8} and G3RS~\cite{tamagaki}.  Both of them contain 
central ($V_{\rm c}$), tensor ($V_{\rm t}$) 
and spin-orbit ($V_{\rm b}$) terms. The ${\vi{L}}^2$ and 
$({\vi L}\cdot{\vi S})^2$ terms of the G3RS potential are ignored.
The binding energy of the $\alpha$ particle is 25.09 for AV8$^{\prime}$
and 25.29\,MeV for G3RS~\cite{fbs}. In what follows we use slightly 
truncated wave functions for the $\alpha$ particle to save computer
time. The truncation is done by ignoring the small component with 
total orbital and spin angular momenta $L=1,\, S=1$ the magnitude 
of which is of the order of 0.3\% or 
by optimizing the wave function  
in a smaller basis set whose size is approximately one third 
of the converged 
solution. The change of the phase shift due to the truncation 
is estimated to be at most a few percent. 

\begin{figure}[t]
\epsfig{file=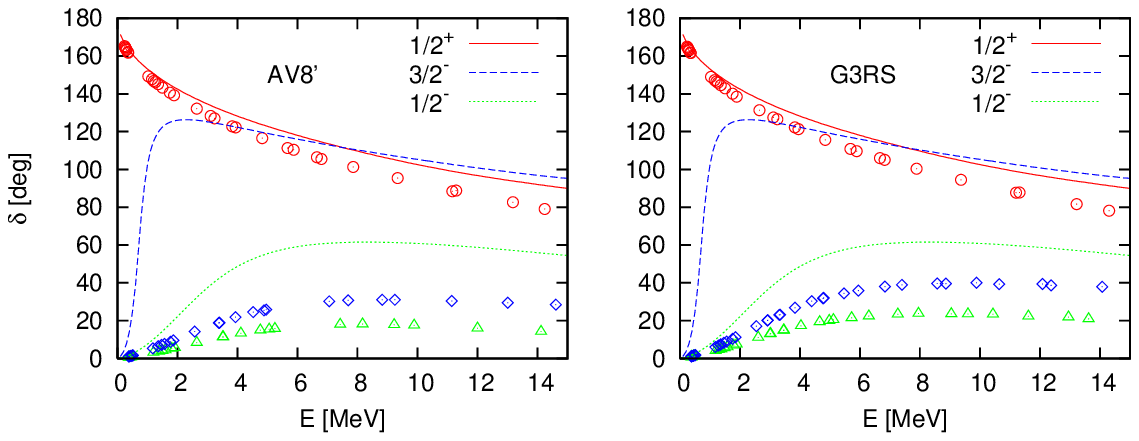,scale=1.3}
\caption{$\alpha$+$n$ scattering phase shifts calculated using 
the AV$8^{\prime}$ and G3RS potentials. Solid, 
dashed and dotted lines are the empirical phase
 shifts~\cite{bond,stamm}. Circle, diamond and triangle symbols denote
the calculated phase shifts for $1/2^+, 3/2^-$ and $1/2^-$, respectively.}
\label{alpha.nreal}
\end{figure}

Figure~\ref{alpha.nreal} displays the $S$- and $P$-wave phase shifts 
obtained in an $\alpha+n$ single-channel calculation. The difference 
due to the nucleon-nucleon potential is not very large. The agreement 
between theory and experiment is quite reasonable for the $1/2^+$ phase
shifts. Our phase shifts agree fairly well with those of the 
Quantum Monte Carlo calculation obtained using 
only the two-body potential of AV18~\cite{nollet}.  
In a sharp contrast to the $S$-wave 
phase shift, the calculated $P$-wave phase shifts are by far smaller than 
experiment, and considerably smaller than those obtained in 
Ref.~\cite{nollet}. We have coupled an inelastic channel of
$^4$He$(0^+_2)+n$ with the elastic channel, but the effect of coupling 
is negligible below $E=$15\,MeV. 

\begin{figure}[b]
\epsfig{file=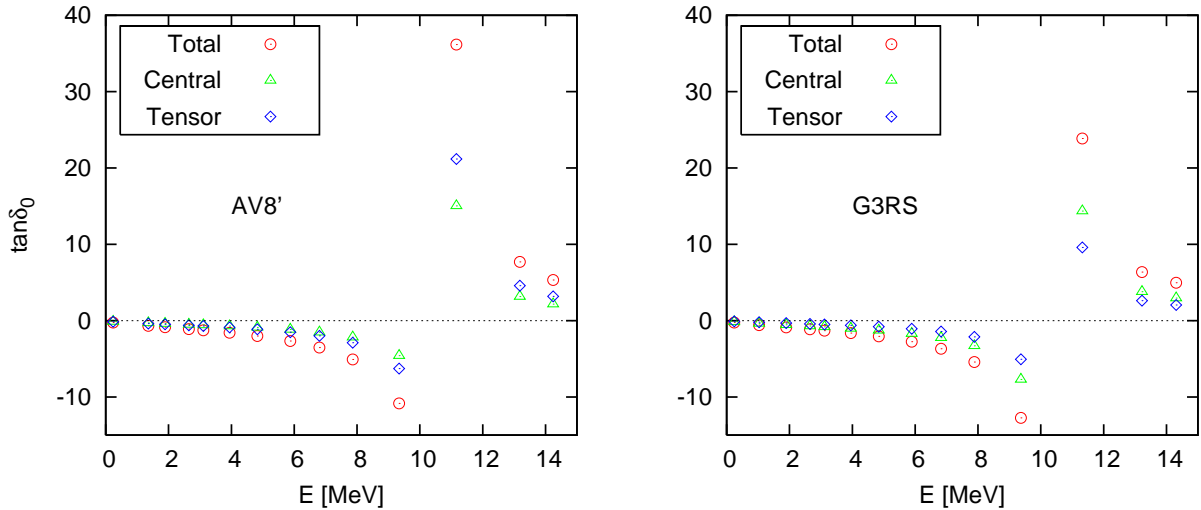,scale=1.3}
\caption{Contributions of the components of the nucleon-nucleon
 potential to $\tan\delta_{0}$ for $S$-wave  
$\alpha$+$n$ scattering phase shifts. The contribution of the spin-orbit
 force is negligibly small.}
\label{tand.comp}
\end{figure}

In Fig.~\ref{tand.comp} we show the contribution of the potential components 
to the $S$-wave phase shift, $\tan \delta_0$. Generally speaking, 
the two 
potentials give similar results. Because the spin-orbit force produces 
a negligible contribution, we do not show its contribution in the figure. 
The phase shift gets largest contributions
from the central and tensor forces. For the AV8$^{\prime}$ potential, the 
tensor contribution is larger than the central contribution, 
whereas, for the G3RS potential, the 
central force gives a larger contribution. This is 
quite consistent with the relative importance of the
tensor and central forces found in the binding energies of $A=3,\,4$ 
nuclei~\cite{fbs}.

The experiment shows a $3/2^-$ sharp resonance around $E=0.9$\,MeV and 
a $1/2^-$ broad resonance. The 
single-channel calculation apparently misses some important 
configurations which are relevant to the resonances, for example 
the inelastic channels including the excited states of $^4$He 
with $J^{\pi}\ne 0^+$, as indicated in Ref.~\cite{quaglioni}. 
Other configurations which could be more important than these 
inelastic channels include the channels of different cluster 
partitions, 
$^3$H+$d$ and $^3$He+$2n$, where $2n$ denotes a di-neutron 
cluster. The threshold energies of these channels are lower than 
those of the inelastic channels. The clusters $^3$H, $d$, and $^3$He may not 
necessarily be in their ground states but can be in their pseudo-excited  
states. This is plausible because the strong tensor component included 
in the realistic potential brings about a mixing of different 
orbital and spin angular momenta of the participating nuclei. 

When we include the different cluster partitions of 
$^3$H+$d$ and $^3$He+$2n$, we have to note 
the old, unresolved problem that the threshold energies of 
different channels are not usually reproduced well. The difference of the 
threshold energies between $\alpha+n$ and $^3$H+$d$ is 17.6\,MeV 
experimentally, but the theoretical value turns out to be 
about 15\,MeV for both the
AV8$^{\prime}$ and G3RS potentials. To quantify the effects of 
the $^3$H+$d$ and $^3$He+$2n$ channels properly, the inconsistency in the threshold 
energy should be settled beforehand. The inclusion of 
three-body forces is important in this respect. 
In the calculation of Ref.~\cite{nollet} including 
the three-body force of the Illinois-2 model, 
the threshold problem does not appear, and some of the distorted 
configurations are certainly included implicitly. 
However, it is not very clear whether the improvement in the phase 
shifts obtained with the use of the three-body force indicates 
simply the predominant importance of the $^3$H$+d$ or 
$^3$He+$2n$ channel or the 
necessity of including further 
configurations other than the elastic channels of 
$\alpha+n$ and $^3$H$+d$. Clarifying this point will 
certainly be very important, but it is beyond the present work because 
such calculations require an expensive computation. 

We here remark on similarity in the phase shifts between $\alpha+n$ 
and $^3$He$+p$ scatterings. The most attractive 
phase shifts of the low-energy $^3$He$+p$ 
scattering occur in the $P$-wave channel with 
$I=1$ and $J=2$, showing a resonance behavior. This is 
consistent with the fact that 
the ground state of $^4$Li has $J^{\pi}=2^-$. 
The $S$-wave phase shift with $I=1$ and 
$J=1$ shows a repulsive behavior similarly to the 
$\alpha+n$ case. Two of the present authors (K. A. and Y. S.) and
S. Aoyama have recently calculated the 
$^3$He$+p$ phase shifts in a microscopic $R$-matrix method 
using the G3RS potential~\cite{arai}. The 
result of the single-channel calculation is very similar to the 
$\alpha+n$ case. The $S$-wave phase shifts are accounted for 
fairly well in the $^3$He$+p$ single-channel calculation, 
but the calculated $P$-wave phase shifts turn out to be too small compared to 
the empirical ones. A calculation of including 
the channels of $d(1^+,0^+)+2p(0^+)$ has 
significantly improved the discrepancy in the phase shifts 
and a further inclusion of 
the inelastic channels of $^3$He$+p$ has reproduced the empirical phase 
shifts reasonably well. We understand this as follows. 
In the $P$-wave scatterings of both $\alpha+n$
and $^3$He$+p$, the nucleon can 
penetrate close to the nucleus and the nucleon-nucleus interaction, 
particularly the tensor force distorts the nuclear state. Thus the 
single-channel assumption becomes rather poor. On the other hand, 
in the $S$-wave case the nucleon cannot come close to the nucleus 
because the two identical nucleons occupying the $S$-orbit repel the 
incoming nucleon due to the Pauli principle, and thus the nucleus 
receives little distortion.

\section{Summary}
\label{summary}

We have formulated a method to calculate the scattering phase shift
using continuum-discretized states. The spectroscopic amplitude
calculated from the discretized state is compared to that 
obtained with the Green's function, 
from which the phase shift can be determined. 
The method has been tested successfully in the cases of 
$n+p$ and $\alpha+n$ scattering where two nucleons are assumed to 
interact via an effective nucleon-nucleon potential. The method has the
advantage that it can be used with ease to scattering problems of
coupled-channels. 

Using a sophisticated wave function for the $\alpha$ particle, we have 
applied the present method to $\alpha+n$ scattering 
described by realistic 
nucleon-nucleon potentials. The $1/2^+$ phase shifts for the 
$S$-wave showing a repulsive behavior of the underlying $\alpha n$ 
interaction are in reasonable agreement with the empirical phase shifts. 
The missing
attraction needed to reproduce the data more perfectly is attributed 
to the effect of three-body forces~\cite{nollet}. We have analyzed 
how much the $S$-wave phase shifts 
are affected by the terms of the nucleon-nucleon 
potential. Both the tensor and central forces contribute significantly 
but the spin-orbit force has a negligible contribution.

The $P$-wave phase
shifts for the $3/2^-$ and $1/2^-$ states are too small 
with respect to 
experiment. This unexpected result indicates that a good reproduction of 
the $\alpha+n$ phase shifts attained using an effective interaction 
such as the Minnesota potential is an {\it ad hoc} description and may not 
be very realistic. Based on the comparison with the result of 
Ref.~\cite{nollet}, the
discrepancy in the $P$-wave phase shifts could be reduced by taking 
into account some distorted configurations which are especially 
important to form the $3/2^-$ resonance as well as the three-body
force. A careful study on the effects of the distorted configurations 
will be important to pin down the contribution of the three-body force.

\bigskip
\vspace{0.5cm}
We would like to thank R. G. Lovas for his careful reading of the
manuscript and useful comments. This work has been
performed as a part of the Bilateral Joint Research Project between 
the JSPS (Japan) and the FNRS (Belgium).  
W. H. is supported by a Grant-in Aid for Scientific Research for 
Young Scientists (No.\,19$\cdot$3978) as a JSPS Research Fellow 
for Young Scientists. 

\appendix

\renewcommand{\theequation}{A\arabic{equation}}
\setcounter{equation}{0}
\section*{APPENDIX: Calculation of a spectroscopic amplitude}
\label{appendix.a}

In this appendix we show a method of calculating $y(r)$ 
for the basis functions used in this article. 
Because the matrix element for 
the spin-isospin parts can be calculated in a standard technique, 
we focus on only the spatial part of the SA. The SA plays an important
role in discussing the spectroscopic properties 
of nuclear structure. 
See Ref.~\cite{inversion} for the case of $^4$He.  
  
Let $({\vi x}_1,{\vi x}_2,\ldots,{\vi x}_{N-2})$ denote the 
internal coordinates of nuclei $\alpha_1$ and $\alpha_2$. Here 
$N$ is the number of nucleons which make up the two nuclei. 
The set of $N-2$ coordinates is simply represented by an $N-2$ 
column vector ${\vi x}$ whose $i$th element is 
${\vi x}_i$. For the sake of convenience, let ${\vi x}_{N-1}$,  
instead of ${\vi r}_c$ used in the main text, denote the relative 
distance vector between the two nuclei. 
Let ${\vi X}$ stand for an $N-1$ column 
vector whose $i$th element is ${\vi x}_i$. The first $N-2$ elements 
of ${\vi X}$ are exactly the same as the elements of ${\vi x}$, and 
the last element of ${\vi X}$ is ${\vi x}_{N-1}$.  

The spatial part of the test function~(\ref{test.fn}) is assumed 
to take the form 
\begin{equation}
\Phi_{LM_L}(r)=\left[F_{(L_3L_4)L'}(u_3,u_4,A',{\vi x})
\frac{\delta(x_{N-1}-r)}{x_{N-1}r}
Y_{\ell}(\hat{\vi x}_{N-1})\right]_{LM_L},
\label{cg}
\end{equation} 
where the function $F$ is the correlated Gaussian basis state whose 
angular part is specified by two global vectors~\cite{fbs}  
\begin{equation}
F_{(L_3L_4)L'M_L'}(u_3,u_4,A',{\vi x})=
{\rm exp}\left(-{\frac{1}{2}}{\widetilde{{\vi x}}} A' {\vi x}\right)
[{\cal Y}_{L_3}({\widetilde{u_3}}{\vi x}) 
{\cal Y}_{L_4}({\widetilde{u_4}}{\vi x})]_{L'M_L'},
\end{equation}
with ${\cal Y}_{\ell m}({\vi r})=r^{\ell}Y_{\ell m}(\hat{\vi r})$. 
Here 
$A'$ is a positive-definite, symmetric matrix of dimension $N-2$, 
and $u_3$ and $u_4$ are $N-2$-dimensional column vectors. To 
simplify the expressions, we use the notation 
${\widetilde{{\vi x}}} A' {\vi x}=\sum_{i,j=1}^{N-2}A'_{ij}{\vi
x}_i\cdot{\vi x}_j$ and 
${\widetilde{u_3}}{\vi x}=\sum_{i=1}^{N-2}{u_3}_i{\vi x}_i$.
The square bracket in Eq.~(\ref{cg}) indicates the angular momentum 
coupling. It will be convenient to rewrite Eq.~(\ref{cg}) as follows:
\begin{eqnarray}
\Phi_{LM_L}(r)=\sum_{M_L'm}
\langle L'M_L' \ell m |LM_L \rangle \int Y_{\ell m}(\hat{\vi r})\, 
F_{(L_3L_4)L'M_L'}(u_3,u_4,A',{\vi x})
\delta({\vi x}_{N-1}-{\vi r}) d{\hat{\vi r}}.
\label{cg.test}
\end{eqnarray}

Similarly, the spatial part for the basis function 
of the composite system, 
$\alpha_1$+$\alpha_2$, is constructed from the function 
\begin{equation}
F_{(L_1L_2)LM_L}(u_1,u_2,A,{\vi X})
={\rm exp}\left(-{\frac{1}{2}}{\widetilde{{\vi X}}} A 
{\vi X}\right)
[{\cal Y}_{L_1}({\widetilde{u_1}}{\vi X}) 
{\cal Y}_{L_2}({\widetilde{u_2}}{\vi X})]_{LM_L}.
\label{cg.compo}
\end{equation}
Here $A$ is a positive-definite, symmetric matrix 
of dimension $N-1$, and $u_1$ and $u_2$ are column vectors of 
dimension $N-1$. The RGM wave function~(\ref{rgm.wf}) 
with Eq.~(\ref{uexp}) can be expressed 
in terms of the above function~(\ref{cg.compo}). That is, 
the matrix $A$ becomes block-diagonal and contains three blocks, 
each of which has the dimension of 
$N_{\alpha_1}-1,\, N_{\alpha_2}-1,\, 1$, respectively. Here 
$N_{\alpha_i}$ is the mass number of the nucleus $\alpha_i$. The  
$A_{N-1\, N-1}$ element corresponds to $\beta_i$. In addition, 
one of the global vectors, say, 
$\widetilde{u_2}{\vi X}$ must be equal to ${\vi x}_{N-1}$ and 
the other 
global vector, $\widetilde{u_1}{\vi X}$, should not contain ${\vi
x}_{N-1}$. This requirement is met simply by choosing 
$u_{2_i}=0\, (i=1,\ldots,N-2)$,  $u_{2_{N-1}}=1$, and 
$u_{1_{N-1}}=0$. We have to note, however, 
that the antisymmetry requirement on the function 
(\ref{cg.compo}) with the spin-isospin parts being included 
is satisfied by redefining $A$, $u_1$ and 
$u_2$ appropriately~\cite{fbs}, and thus the above-mentioned 
simplicity in $A$, 
$u_1$ and $u_2$ is destroyed when implementing the permutational 
symmetry.

The SA is obtained through the overlap integral of 
Eqs.~(\ref{cg.test}) 
and (\ref{cg.compo}). The integration extends over all the elements 
of ${\vi X}$. Because of the function 
$\delta({\vi x}_{N-1}-{\vi r})$, the 
coordinate ${\vi x}_{N-1}$ contained 
in Eq.~(\ref{cg.compo}) may be replaced 
by ${\vi r}$. This replacement can be done as follows. 
By decomposing the matrix $A$ into 
\begin{equation}
A
=\left(
\begin{array}{cc}
A^{(1)} & a^{(1)} \\
\widetilde{a^{(1)}} & a \\
\end{array}
\right),
\label{def.A}
\end{equation}
the exponent of Eq.~(\ref{cg.compo}) is reduced to the form
\begin{equation}
{\widetilde{{\vi X}}} A {\vi X}={\widetilde{{\vi x}}} A^{(1)} {\vi x}
+2\widetilde{a^{(1)}}{\vi x}\cdot{\vi r}+ar^2.
\end{equation}
Here the matrix $A^{(1)}$ is symmetric and has dimension $N-2$, 
the column vector $a^{(1)}$ is also of dimension $N-2$, and $a$ is equal
to $A_{N-1\,N-1}$. 
The vector of the angular part is reduced, e.g., to 
\begin{equation} 
{\widetilde{u_1}}{\vi X}={\widetilde{u_1^{(1)}}}{\vi x}+
u_{1_{N-1}}{\vi r},
\end{equation}
where $u_1^{(1)}$ is a column vector which consists 
of the first $N-2$ elements of $u_1$.

Now the overlap integral for the SA reads
\begin{eqnarray}
& &\langle \Phi_{LM_L}(r)|F_{(L_1L_2)LM_L}(u_1,u_2,A,{\vi X})\rangle
\nonumber \\
&& =\sum_{M_L'm}
\langle L'M_L' \ell m |LM_L \rangle \int Y_{\ell m}^*(\hat{\vi r})\,
{\rm e}^{-\frac{1}{2}ar^2}
 \Big\langle 
{\rm exp}\left(-{\frac{1}{2}}{\widetilde{{\vi x}}} A' {\vi x}\right)
[{\cal Y}_{L_3}({\widetilde{u_3}}{\vi x}) 
{\cal Y}_{L_4}({\widetilde{u_4}}{\vi x})]_{L'M_L'}\Big|
\nonumber \\
&& \times \, 
{\rm exp}\left(-{\frac{1}{2}}{\widetilde{{\vi x}}} A^{(1)} {\vi x}
-\widetilde{a^{(1)}}{\vi x}\cdot{\vi r}\right)
[{\cal Y}_{L_1}({\widetilde{u_1^{(1)}}}{\vi x}+
u_{1_{N-1}}{\vi r}) 
{\cal Y}_{L_2}({\widetilde{u_2^{(1)}}}{\vi x}+
u_{2_{N-1}}{\vi r})]_{LM_L}\Big\rangle d{\hat{\vi r}}.
\end{eqnarray}
With the change of variables from ${\vi x}$ to ${\vi t}$,  
${\vi x}$=${\vi t}$+$\omega {\vi r}$, where 
\begin{equation}
\omega=-B^{-1}a^{(1)},\ \ \ \ \ B=A^{(1)}+A',
\end{equation}
the overlap integral becomes 
\begin{eqnarray}
&&\langle \Phi_{LM_L}(r)|F_{(L_1L_2)LM_L}(u_1,u_2,A,{\vi X})\rangle
\nonumber \\
& &=\sum_{M_L'm}
\langle L'M_L' \ell m |LM_L \rangle \int Y_{\ell m}^*(\hat{\vi r})\,
{\rm e}^{-\frac{1}{2}(a+\widetilde{a^{(1)}}\omega)r^2}
{\cal M}({\vi r})d{\hat{\vi r}},
\label{SA.int}
\end{eqnarray}
where 
\begin{eqnarray}
{\cal M}({\vi r})&=&\langle 
{\rm e}^{-\frac{1}{2}\widetilde{\vis t}A'{\vis t}}
[{\cal Y}_{L_3}({\widetilde{u_3}}{\vi t}+b_3{\vi r}) 
{\cal Y}_{L_4}({\widetilde{u_4}}{\vi t}+b_4{\vi r})]_{L'M_L'}|
\nonumber \\
&\times&\, 
{\rm e}^{-\frac{1}{2}\widetilde{\vis t}A^{(1)}{\vis t}}
[{\cal Y}_{L_1}({\widetilde{u_1^{(1)}}}{\vi t}+b_1{\vi r}) 
{\cal Y}_{L_2}({\widetilde{u_2^{(1)}}}{\vi t}+b_2{\vi r})]_{LM_L}
\rangle,
\end{eqnarray}
with
\begin{equation}
b_1=\widetilde{u_1^{(1)}}\omega+u_{1_{N-1}},\ \ \ 
b_2=\widetilde{u_2^{(1)}}\omega+u_{2_{N-1}},\ \ \ 
b_3=\widetilde{u_3}\omega,\ \ \ 
b_4=\widetilde{u_4}\omega.
\end{equation}

To obtain the SA, we have to carry out the integration over 
$\hat{\vi r}$ in 
Eq.~(\ref{SA.int}). 
The dependence of ${\cal M}({\vi r})$ on ${\hat{\vi r}}$ is 
extracted using a decomposition of the type of 
\begin{equation}
{\cal Y}_{L_1}({\widetilde{u_1^{(1)}}}{\vi t}+b_1{\vi r})
=\sum_{\ell_1=0}^{L_1}D^{L_1}_{\ell_1}
{\cal Y}_{L_1-\ell_1}({\widetilde{u_1^{(1)}}}{\vi t})
{\cal Y}_{\ell_1}(b_1{\vi r})
\end{equation}
with
\begin{equation}
D^{L}_{\ell}=\sqrt{\frac{4\pi(2L+1)!}{(2\ell+1)!(2L-2\ell+1)!}},
\end{equation}
and recoupling the angular momenta which come from the four 
vectors $b_i{\vi r}$. The bra-ket functions depending on ${\vi t}$ 
lead to the overlap matrix element of the $N-2$-particle system.
After these algebraic manipulations, we get the desired formula 
\begin{eqnarray}
& &\langle \Phi_{LM_L}(r)|F_{(L_1L_2)LM_L}(u_1,u_2,A,{\vi X})\rangle
\nonumber \\
&& =\sum_{\ell_1 \ell_2 \ell_3 \ell_4}
b_1^{\ell_1}b_2^{\ell_2}b_3^{\ell_3}b_4^{\ell_4}\,
D^{L_1}_{\ell_1}D^{L_2}_{\ell_2}D^{L_3}_{\ell_3}D^{L_4}_{\ell_4}\,
r^{\ell_1+\ell_2+\ell_3+\ell_4}\,
{\rm e}^{-\frac{1}{2}(a+\widetilde{a^{(1)}}\omega)r^2}
\nonumber \\
&& \times \sum_{\ell_{12}\ell_{34}\Lambda}
C(\ell_1 \ell_2;\ell_{12})C(\ell_3 \ell_4;\ell_{34})
C(\ell \ell_{34};\ell_{12})U(\Lambda \ell_{34} L \ell;L' \ell_{12})
\nonumber \\
&& \times 
\left[
\begin{array}{ccc}
L_1-\ell_1 &  \ell_1    & L_1  \\
L_2-\ell_2 &  \ell_2    & L_2  \\
\Lambda    &  \ell_{12} & L \\
\end{array}\right]
\left[
\begin{array}{ccc}
L_3-\ell_3 &  \ell_3    & L_3  \\
L_4-\ell_4 &  \ell_4    & L_4  \\
\Lambda    &  \ell_{34} & L' \\
\end{array}\right]
\nonumber \\
&&\times 
\langle F_{(L_3-\ell_3\, L_4-\ell_4)\Lambda M_{\Lambda}}(u_3,u_4,A',{\vi x})|
F_{(L_1-\ell_1\, L_2-\ell_2)\Lambda M_{\Lambda}}
(u_1^{(1)},u_2^{(2)},A^{(1)},{\vi x})\rangle,
\label{grandformula}
\end{eqnarray}
where $C$ is a coefficient which couples two spherical harmonics 
with the same argument
\begin{equation}
C(l_1l_2;l_{12})=\sqrt{\frac{(2l_1+1)(2l_2+1)}{4\pi (2l_{12}+1)}}
\langle l_10l_20\vert l_{12}0 \rangle,
\end{equation}
$U$ is a unitary Racah coefficient, and 
the square bracket $\left[ \cdots \right]$ stands for 
a unitary 9-$j$ coefficient~\cite{svm}. The overlap matrix 
element in the last line of Eq.~(\ref{grandformula}) can be 
calculated using Eq.~(B.10) of Ref.~\cite{fbs}.

\end{document}